# The Anderson-Mott transition induced by hole-doping in Nd$_{1-x}$TiO$_3$

A.S. Sefat, J.E. Greedan

*Departments of Chemistry, McMaster University, Hamilton, Ontario, Canada*

G.M. Luke[a], M. Niewczas[b]

*Department of Physics and Astronomy [a] and Materials Sciences and Engineering [b], McMaster University, Hamilton, Ontario, Canada*

J.D. Garrett, H. Dabkowska, A. Dabkowski

*Brockhouse Institute for Materials Research, McMaster University, Hamilton, Ontario, Canada*

The insulator/metal transition induced by hole-doping due to neodymium vacancies of the Mott- Hubbard antiferromagnetic insulator, Nd$_{1-x}$TiO$_3$, is studied over the composition range $0.010(6) \leq x \leq 0.243(10)$. Insulating *p*-types conduction is found for $x \leq 0.071(10)$. Anderson localization in the presence of a Mott-Hubbard gap, is the dominant localization mechanism for the range of $0.074(10) \leq x < 0.089(1)$ samples. For $x \geq 0.089(1)$, *n*-type conduction is observed and the activation energy extrapolates to zero by $x \cong 0.1$. The $0.095(8) \leq x < 0.203(10)$ samples are Fermi-liquid metals and the effects of strong electronic correlations are evident near the metal-to-insulator boundaries in features such as large Fermi liquid $T^2$ coefficients. For $0.074(9) \leq x \leq 0.112(4)$, a weak negative magnetoresistance is found below ~ 15 K and it is attributed to the interaction of conduction electrons with Nd$^{3+}$ magnetic moments. Combining information from our companion study of the magnetic properties of Nd$_{1-x}$TiO$_3$ solid solution, a phase diagram is proposed. The main conclusions are that long range antiferromagnetic order disappears before the onset of metallic behavior and that the Anderson-Mott transition occurs over a finite range of doping levels. Our results differ from conclusions drawn from a similar study on the hole doped Nd$_{1-x}$Ca$_x$TiO$_3$ system which found the co-existence of antiferromagnetic order and metallic behavior and that the Mott transition occurs at a discrete doping level.

## I. INTRODUCTION

One of the most fascinating branches of condensed matter science is the investigation of strongly correlated electron systems. The discovery of high-$T_c$ superconductors has given a huge impetus to the study of correlation effects. Most of the pure, parent materials of the high-$T_c$ superconductors are antiferromagnetic insulators and electronic conduction is established by doping with, for example, $Sr^{2+}$ in $La_{2-x}Sr_xCuO_4$ (LASCO), $O^{2-}$ in $YBa_2Cu_3O_{6+y}$ (YBCO) or $Ce^{4+}$ in $Nd_{2-x}Ce_xCuO_4$. For each of these systems, the Néel temperature decreases by doping and is followed by a crossover to the superconducting phase through a spin-glass phase. For LASCO, the superconducting transition temperature has the maximum $T_c \sim 40$ K for $x = 0.15$, while for YBCO, $T_c \sim 90$ K at $y = 0.7$, and for $Nd_{2-x}Ce_xCuO_4$, $T_c \sim 25$ K around $x = 0.15$ [1]. The renewed interest in $d^1$ transition metal oxide compounds is inspired as such materials are one-electron analogues for the cuprate superconductors with $^2D$ $Cu^{2+}$ free ion ground state ($S = 1/2$, $d^9$). Colossal magnetoresistance in manganites and charge-ordering phases in nickelates are some further spectacular examples of the effects of doping induced electronic phase transitions [2].

The focus of this study is the 3$d$ band-filling of the mixed-valent $Nd_{1-x}TiO_3$ (NTO) system and the associated electronic transitions. For compounds of transition metals at the beginning of the series, such as the titanates, the Mott excitation gap is formed between the top of the lower Mott-Hubbard band (LHB) with single occupancy of the atomic sites and the bottom of the upper Mott-Hubbard band (UHB), separated from the former by a gap, $E_g$. In the perovskite titanates, vanadates, chromates, with the octahedral metal environments, the $t_{2g}$ levels are occupied. Because the $t_{2g}$ orbitals point away from the 2$p$ orbitals, the hybridization with oxygen ligands is very weak and the bandwidth ($W$) is small. Also, the $d$-orbitals are much higher in energy than the 2$p$ orbitals of oxygen. As a result, the charge-transfer energy ($\Delta$), which is the difference between the relative position of oxygen ($\varepsilon_p$) and transition-metal ($\varepsilon_d$) levels, is large. The local *d-d* Coulomb electron repulsion (*U*) is smaller than Δ but larger than the bandwidth. This leads to Mott-Hubbard insulators (MI) in the Zaanen-Sawatsky-Allen (ZSA) scheme [3] for which carriers are localized.

Earlier, we have explored the effect of doping on the electronic structure by optical absorption studies [4]. Figure 1 represents the real part of optical conductivity at room temperature for the parent $d^1$ $Nd_{0.981(6)}TiO_3$ Mott-Hubbard insulator. This spectrum consists of two separate electronic-absorption transitions. For the $Nd_{0.981(6)}TiO_3$ compound with $x = 0.019(6)$, the charge-transfer gap due to O(2*p*)-Ti(3*d*) transition is ~ 4.0 eV, and the Mott-Hubbard gap ($E_g \sim U-W$) is evident at 0.8 eV. The Mott-Hubbard gap is consistent with those reported for $LaTiO_3$ and $YTiO_3$ of 0.2 eV and 1 eV, respectively [5]. The relative positions of the two gap excitations indicate that the parent $d^1$ $Nd_{0.981(6)}TiO_3$ is a Mott insulator. The intra-atomic Coulomb repulsion energy (*U*) of 4.0 eV was found for $LaTiO_3$ and $La_{0.90}Sr_{0.10}TiO_3$ using resonant soft-x-ray emission (SXES) in combination of photoemission and inverse-photoemission spectroscopy (PES/IPES) [6]. Also, *U* has been viewed as a constant among the $RTiO_3$ series [7]. Thus, we estimated the magnitude of ~ 3.2 eV for the full width of the one-electron LHB of $NdTiO_3$ [4]. This value is an estimation, and larger than the calculated $W = 2.27$ eV value reported for $NdTiO_3$ using the tight-binding model [7].

The mixed-valent $Nd_{1-x}TiO_3$ can be considered as a solid solution between the two perovskite related end members of $NdTiO_3$ ($x = 0$, $Ti^{3+}$, $3d^1$) and $Nd_{2/3}TiO_3$ ($x = 0.33$, $Ti^{4+}$, $3d^0$), a band insulator. The electronic structure of these two materials are shown, schematically in Figure 2 labeled A and D. Substitution of neodymium vacancies in $NdTiO_3$ decreases the number of electrons per titanium site from 1 at a rate of 3*x*. At a critical concentration of electrons, an electronic transition to a metal is expected, called the Mott transition (MIT1). While the two end members are not strictly isostructural, $NdTiO_3$ is described in *Pnma* [8] and $Nd_{0.67}TiO_3$ in *Cmmm* [9], the structures are closely related as the principal difference being that the Nd vacancies are crystallographically ordered in *Cmmm*. Upon hole doping, the *Pnma* form persists for $0.0 < x < 0.25$ and the *Cmmm* for $x = 0.30$ to 0.33 [9] (Figure 3). For low doping levels, this has consequences for the electronic structure as shown in Figure 2, section B. New states will be added with energies within the Mott Hubbard gap but near the top of the LHB [3]. At low doping levels the carriers in the LHB would be holes. With further hole doping, the band width of these mid-gap states would increase and might at some point overlap with the top of the LHB, resulting in a metallic-like state but without complete collapse of the Mott Hubbard gap ($E_g$). This transition is called the Mott transition, indicated as the first metal-to-insulator transition (MIT1) on Figure 2. With further doping and the associated increase in bandwidth, the Mott-Hubbard gap would collapse giving a more typical metal within the ZSA picture as shown in Figure 2, labeled C. A second insulator-to-metal transition (MIT2), section D in Figure 2, will result upon addition of neodymium ions to $Nd_{0.67}TiO_3$, by introducing carriers, in this case electrons, at a critical carrier density of $x \sim 0.20$ [8]. In previous work on the NTO system, the metallic region has been roughly demarcated for the doping range $0.10 \leq x \leq 0.20$ [8]. In our present study, we locate the first electronic transition more precisely by finely controlled filling of the lower Mott-Hubbard band through vacancy-doping. The doping range investigated is illustrated by the arrows on the phase diagram in Figure 2. It has been of critical importance to measure the vacancy concentration, $x$, in $Nd_{1-x}TiO_3$. Details of the compositional analyses have been described elsewhere [10]. In brief, neutron activation and thermal gravimetric analyses were used along with precise measurements of the unit cell volumes using x-ray powder diffraction. From the neutron activation technique, the Nd/Ti ratio was determined by comparing to reference samples with known Nd/Ti ratios. Also, the oxidative weight gain of each sample was monitored to verify the $Ti^{3+}$ present per unit formula.

The metal-insulator transitions (MIT) in oxides have been studied extensively. There are various scenarios that can cause such electronic transitions. Among them are carrier densities and bandwidth changes (as a result of bond angle change for example), electron correlation and also disorder. Inclusion of disorder may result in the presence of mobility edges, which is a further carrier localization mechanism. Some well-studied systems include $SrTi_{1-x}Ru_xO_3$, with six distinct electronic states and a MIT at $x \sim 0.5$ [11], and $CaV_{1-x}Ti_xO_3$, with MIT for $0.2 < x < 0.4$ [12]. However, these systems are qualitatively different in that the MIT involves a correlated metal and a band insulator, whereas, $NdTiO_3$ is a Mott Hubbard insulator. For this study, numerous $Nd_{1-x}TiO_3$ polycrystalline samples as well as five single crystals with $x \sim 0, 0.04, 0.10, 0.15$, and 0.20 have been investigated. In the mixed-valent system $Nd_{1-x}Ti_{1-3x}^{3+}Ti_{3x}^{4+}O_3$, the doping is associated with the Nd-sublattice and the disorder potential of the randomly distributed vacancies is expected to be relatively strong. The temperature dependence of the electronic resistivity ($\rho$), magnetoresistance and thermopower ($S$) are reported for the compositions. Fine control of the doping levels has disclosed details of the Anderson-Mott transition, not reported in previous studies on the $Nd_{1-x}TiO_3$ system nor in related materials such as $Nd_{1-x}Ca_xTiO_3$ [13]. In the $Nd_{1-x}Ca_xTiO_3$ study, the coexistence of antiferromagnetic order and metallic behavior was reported and the Mott transition was found to occur at a discrete doping level. It is of considerable interest to compare the results from the closely related studies.

## II. EXPERIMENTS

$Nd_{1-x}TiO_3$ (NTO) samples with $0.010(6) \leq x \leq 0.243(10)$ were prepared by mixing, grinding and pelletizing stoichiometric amounts of $Ti_2O_3$ (Cerac, 99.9%), pre-dried $Nd_2O_3$ (Research Chemicals 99.99%) and $TiO_2$ (Fisher Scientific 99.97%). Each sample was sealed in a molybdenum crucible under purified argon gas. The preparation conditions for polycrystalline samples involved several firing steps, at ~1400 °C for ~12 hours in an *rf* induction furnace. Single crystal growth using the Bridgeman method for samples of $x \sim 0.04$ and $x \sim 0.10$ was performed by uniform melting of the mixed raw materials in a welded molybdenum crucible and slow cooling through a temperature-gradient in the induction furnace. The single crystal growth for the $x = 0$, 0.15 and 0.20 NTO samples was carried out by a floating zone (FZ) technique using a double ellipsoid image furnace (NEC SCI-MDH-11020). For each individual run, ~20 g of mixed powders was moulded into two cylindrical rods. These, the feed and seed rods, were made by packing the powder evenly in a rubber tube and sealing and pressing in a hydrostatic press. The radii of the rods were ~5 mm and the lengths were variable. For these growths, the translational velocity of the seed rod was ~ 25 mm hour$^{-1}$ and the experiments were performed in ~ 3.5 atm of flowing 5% $H_2$/Ar gas mixtures.

Phase purity of the samples was initially monitored by x-ray powder diffraction using a Guinier-Hägg camera with monochromated Cu K$\alpha_1$ radiation and a silicon standard. Neutron activation analyses (NAA) of the sintered samples were used to fix the vacancy levels, by comparison of the neodymium to titanium ratio to those of reference samples. Having a measure of the Nd/Ti value from neutron activation measurements, the thermogravimetric (TGA) weight gain of each $Nd_{1-x}TiO_3$ sample, oxidized to $Nd_2Ti_2O_7$ and $TiO_2$, was monitored in flowing air at 1000°C. The results from NAA and TGA were consistent to ~ 0.5%. The NAA experiments were performed at the McMaster Nuclear Reactor (MNR) and the TGA was done using a 409-Netzch or PC-Luxx Simultaneous Thermal Analyzer.

The electrical resistivity of all of the prepared single crystal and polycrystalline NTO samples was measured using a Quantum Design Physical Property Measurement System (PPMS) using a four-probe geometry. For preparing the samples for measurements, blocks of single crystal and polycrystalline samples were cut into bar-shapes using a diamond saw. Onto each sample, 4 gold contacts were sputtered using a Sputter Coater S150B in 0.3 mbar atmosphere of argon. Then, silver adhesive paste (Alfa Aesar) was used to attach silver wires. Each sample was measured in the range of 2 K to 300 K using a constant excitation current of 0.1 mA or 1 mA, depending on sample resistance. For the AC runs, a frequency of 133 Hz was employed. A typical sample size was ~ 7 mm (length) x 2 mm (thickness) x 3 mm (width). In DC mode, at each temperature point, 25 consecutive voltage readings were averaged; each sample was measured in the range of 2 K to 300 K in zero field and a magnetic field of 8 T. In addition, field scans in the range 0 T to 9 T were performed at specific temperature points.

The thermoelectric power measurements were done on several compositions of NTO by using a local apparatus [8]. The procedure for measuring the Seebeck coefficient involved the steady-state straddle method. The Seebeck voltage was measured, from two copper wires attached to the sample with silver paste directly across from the thermocouples. The gradient was created across the samples by controlling the temperature of head A at $T_1$ and varying the temperature of head B over the range $T_1 \pm 5$ K in 1 K steps. For each temperature step, 15 data points were taken for statistical averaging.

Neutron Powder Diffraction data were collected on the C2 diffractometer operated by the Neutron Program for Materials Research of the National Research Council of Canada at Chalk River Laboratories of AECL Ltd. The samples of ~ 1- 2 g were placed in helium-filled vanadium cans that were sealed with an indium gasket. The chemical structural refinements for several vacancy doped samples were obtained from neutron powder data ($\lambda \cong 1.32$ Å) collected at room temperature. For the magnetic structure determinations, data were collected at ~4 K using ($\lambda \cong 2.37$ Å) neutrons. For more details & descriptive results of such experiments, reference [10] may be used.

### III. STRUCTURAL CHANGES UPON HOLE-DOPING

Before discussing the changes in electrical transport properties due to hole-doping of $NdTiO_3$, the structural changes, determined by neutron powder diffraction and reported in a companion paper, should be reviewed [10]. These results are best summarized by Figure 4. Clearly, two important trends which have a bearing on the electronic structure are apparent. First, the average Ti – O – Ti bond angle increases upon hole-doping from $149.0(4)°$ at $x = 0.019(6)$ to $153.3(4)°$ at $x = 0.098(10)$. Secondly, both the average Ti – O distance and the distortion of the Ti – O octahedron (difference between the two Ti – O bond lengths) decrease. Collectively, from these observations, one expects the width of the lower Hubbard band, $W$, to increase with hole-doping which will lead to a systematic decrease in the Mott-Hubbard gap. An attempt can be made to quantify this effect, at least roughly, by comparisons with the optical data on the $La_{1-x}Y_xTiO_3$ system [5]. It is well known that the average Ti – O – Ti angle is controlled largely by the size of the rare earth ion in the undoped $RTiO_3$ phases [14]. We note that the $La_{0.6}Y_{0.4}TiO_3$ has an average $R$-site radius equivalent to that of $Nd^{3+}$ and encouragingly, the Mott-Hubbard gap found for this composition by is ~ 0.8 eV, the same value as for our report on $NdTiO_3$ [4]. The gap decreases with increasing average $R$-site radius to 0.2 eV for $LaTiO_3$. The increase in Ti – O – Ti angle upon hole doping in $Nd_{1-x}TiO_3$ over the range $0.019(6) < x < 0.098(10)$ is $4.3(6)°$. This is slightly smaller that the increase in proceeding from $NdTiO_3$ to $LaTiO_3$, $5.35(13)°$ [14]. Thus, one might expect a gap decrease of ~ 0.5– 0.6 eV due to the structural changes alone which accompany the hole-doping of NdTiO$_3$. Clearly, the Mott-Hubbard gap will not collapse solely by this effect and it will be reduced by roughly a factor of two over the range of hole-doping which includes the Mott transition.

## IV. ELECTRICAL RESISTIVITY

The results of the temperature dependence of the electrical resistivity, for $x$ spanning the Mott transition (MIT1) in the mixed-valent Nd$_{1-x}$TiO$_3$, are shown in Figure 5. As evident, there are smooth transitions with $x$ from insulating-to-metallic states. Previous electrical resistivity measurements had shown semiconducting behavior for $x = 0$ and $x = 0.05$, and metallic behavior for $x = 0.10$ NTO sample [8]. Here, by finely controlled hole-doping of NTO, the complexity of the Mott transition is apparent. Semiconducting behavior, at least over some significant temperature range, is clearly seen for samples with $x \leq 0.07(10)$ and fully metallic behavior for $x \geq 0.095(8)$. For the intermediate regime, $0.074(9) \leq x \leq 0.089(1)$, samples which span the Mott transition, more complex behavior is observed.

These resistivity results can be compared with the behavior of thermoelectric power measured on a smaller number of samples (Figure 6a). The thermopower of Nd$_{1-x}$TiO$_3$ has been reported previously for $x = 0$ and $x = 0.05$ for which $p$-type conduction has been observed, whereas, $n$-type behavior was seen for the more heavily doped compositions of $x \geq 0.10$ [8]. In the present study, we find that the thermopower changes sign from $p$-type for $x \leq 0.071(10)$ to $n$-type at $x > 0.089(1)$. However, for $x = 0.071(10)$, there is a cross-over to a slightly negative value near room temperature, suggesting a two carrier model. These results can be compared to those reported for La$_{1-x}$Sr$_x$TiO$_3$ (Figure 6b) in which the thermopower changes abruptly from $p$-type for $x \leq 0.044$ to $n$-type at $x = 0.05$ [15] with no apparent evidence for a two carrier regime. Note that the $p$-type to $n$-type transition in La$_{1-x}$Sr$_x$TiO$_3$ occurs at a carrier concentration of ~ 5% holes, whereas that for Nd$_{1-x}$TiO$_3$ requires a much greater nominal hole concentration of ~ 21%. A more detailed analysis of the NTO sample results for various composition ranges follows this discussion.

## IV.A. Semiconductors and samples spanning MIT1

### IV.A.I Semiconductors: $0.010(6) \leq x \leq 0.071(10)$

For $Nd_{1-x}TiO_3$, samples with $0.010(6) \leq x \leq 0.071(10)$ in region I of the phase diagram (Figure 2), semiconducting behavior is seen over a significant temperature range in Figure 5. For the small doping levels of $x = 0.010(6)$ and $0.046(10)$, semiconducting behavior is observed over the entire temperature range studied and the schematic of the density of states is likely close to that pictured as 'A' in Figure 2. However, especially for $x = 0.046(10)$, the involvement of mid gap states will be significant. The thermopower results from Figure 6a indicate nearly temperature independent behavior above 170 K for the lightly doped $x = 0.019(6)$ which supports a polaronic hopping conduction mechanism involving $p$-type carriers.

For $x = 0.057(11)$ and $0.071(10)$, ρ decreases with increasing temperature up to ~130 K, above which it displays nearly temperature-independent behavior with a slight upturn. Evidently, the character of the charge carriers for these samples seems to change from polaronic towards an itinerant state at ~ 130 K. As discussed earlier, the crystallographic changes upon hole doping of $NdTiO_3$ will result in a broadening of the LHB and with increasing hole doping the mid-gap states bandwidth will also increase. Thus, the Fermi level will move closer to the mid-gap band, as depicted by the 'B' density-of-states in Figure 2 which will result in a decrease in the activation energy for polaronic hopping, $E_{act}$, to levels comparable to thermal energies, $k_BT$, in this range [4]. Direct evidence for this will be presented in a following section. This picture is qualitatively consistent with the broad maxima in the thermopower vs temperature curves for $x = 0.064(10)$ and $0.071(10)$ noted earlier in Figure 6a. For these samples, the Mott-Hubbard gap is likely still > 0.2 eV and it is the mid-gap states which are in play here rather than the UHB. Carriers in these states should be $n$-type and at ~300 K, if the contributions of holes in the LHB and electrons in the mid-gap band bands are comparable then the total thermopower would approach zero and could even become slightly negative above room temperature, as seen in Figure 6a.

**IV.A.II   Samples spanning MIT1:  0.074(9) < $x$ < 0.089(1)**

The resistivity curves, $\rho(T)$ in Figure 5, for $x$ = 0.074(9), 0.079(2), and 0.089(1) all show smooth transitions near 150 K from semiconductive to metallic temperature dependences upon heating from 2 K. Before considering the origin of this transition, it is useful to examine in more detail the semiconducting behavior of all of the materials in this class. The temperature dependence of resistivity for the samples in the range 0.010(6) ≤ $x$ ≤ 0.089(1), have been fitted to a simple Arrhenius law, $\rho = \exp(-\Delta_{act}/k_B T)$, as shown in Figure 7a. The derived activation energies ($\Delta_{act}$) are listed in Table 1. The value for the most lightly doped sample, e.g. $x$ = 0.010(6), is 0.084 eV, comparable to the 0.08 eV reported for LaTiO$_3$ [15] and 0.086 eV for CeTiO$_3$ [16] which are all in the expected range for small polaron hopping in transition metal oxides. Because these activation energies are an order of magnitude smaller than the measured Mott-Hubbard gap of 0.8 eV, strong evidence that excitations to the mid-gap states, rather than to the UHB, dominate the conductivity in the temperature range studied here. From Figure 7b, $\Delta_{act}$ decreases sharply with increasing $x$, reaching values of only ~ 200 K for $x$ = 0.057(11) and 0.071(10), supporting the arguments made earlier for these compositions to explain the temperature independent resistivity found above 150 K. Finally, $\Delta_{act}$ appears to vanish for $x$ = 0.098(10). It is not clear from the transport data alone whether this onset of metallic behavior is due to overlap between the LHB and the mid-gap band or the actual collapse of the MH gap, *i.e.* the overlap of the LHB and the UHB. Measurements of the optical conductivity indicates that a small gap may still exist between the top of the LHB and the bottom of the mid-gap band for $x$ = 0.095(8), since a midinfrared peak is observed in addition to the Drude component [4].

An attempt was made to define the metal-insulator boundary, more precisely, by analysis of the low temperature conductivity data as suggested by Möbius [17], and applied for example in LaNi$_{1-x}$Co$_x$O$_3$ to find the composition for the MIT at $x \cong 0.65$ [18]. In this approach it is assumed that the conductivity at sufficiently low temperatures is described by $\sigma(T) = \sigma(0) + CT^z$ and the quantity $w(T) = d\ln\sigma / d\ln T$ is plotted versus T. If the sample is metallic and shows finite conductivity at absolute zero, $w(T) = zCT^z / [\sigma(0) + CT^z] = zCT^z/\sigma(T)$, and should vanish at absolute zero. On the other hand, weakly insulating samples may exhibit temperature independent values of *w*. Unfortunately, application of this method for samples in the composition range $0.074(9) < x < 0.089(1)$, yielded no clear result in part since our data extend only to 2 K and the extrapolation to the mK region is uncertain. In addition, it is clear that the power law asssumed by Möbius is not the only component to the low temperature conductivity in the NTO materials, given the demonstration of a weak magnetoresistance at temperatures below 15 K for samples near the MIT boundary, as it will be demonstrated later.

### IV.A.III   Evidence for the Role of Disorder: Variable-Range Hopping

The process of hole-doping will nearly always involve the introduction of disorder in the material being doped. Given the high doping levels used here, of order of several percent, one expects a significant role for disorder. In disordered solids, the random placement of vacancies gives rise to a random component to the potential seen by the nominally itinerant charge carriers and this can cause localization. Such a disorder induced metal-insulator transition, is known as the Anderson transition [19]. The random potential results in a distribution of activation energies which is manifested experimentally, at sufficiently low temperatures, by the phenomenon of variable range hopping (VRH) which takes the form $\sigma = \sigma_o \exp[-(T_o/T)^y]$, where y is the hopping index. For three spatial dimensions and ignoring correlation effects, $y = ¼$ while the inclusion of correlation gives $y = ½$ [20]. The confirmation of this form of conduction is provided by the linearity of log σ vs $T^{-y}$. The exponent, y, is best extracted from the conductivity results by analyzing the data using the expression [20]:

$$E = \left(\frac{-1}{T}\right)\left(\frac{\partial \log \sigma}{\partial (T^{-1})}\right).$$

The dependence of logE vs. logT for $Nd_{1-x}TiO_3$ samples in the doping range $0.057(11) \leq x \leq 0.095(8)$ is shown in Figure 8a. Note the clear presence of a linear regime for all samples in the temperature range of 27 to 63 K. The y exponents derived are summarized in Figure 8b. For the $0.057 \leq x \leq 0.079(2)$ samples, exponents of $y \cong 0.5$ were found, indicating the presence of correlation effects. Note that y decreases systematically with increasing *x*, reaching a value near 0.25 for $x = 0.089(1)$, which indicates a diminished role for correlation with increasing disorder levels. As a result, for y = 0.5, $E_f \leq E_c$ and so VRH occurs near the mobility edge in the fitted temperature region [21]. Therefore, Anderson localization is the dominant carrier localization mechanism for $0.057 \leq x \leq 0.089(1)$ in the temperature range 27 K to 63 K.

It is expected that the range over which carriers can hop will decrease with increasing temperature until nearest neighbor hopping is the dominant mode of transport and thus, VRH will give way to a simple Arrhenius mechanism. In these samples, transitions to an Arrhenius activation process were found above 63 K as shown in Figure 7. To illustrate more clearly the interplay between VRH, Arrhenius and itinerant conduction processes for a typical sample in this very interesting intermediate doping regime, the temperature dependence for the $x = 0.079(2)$ is shown in Figure 9.

Finally, at this point it is relevant to make a connection with the magnetic properties reported by us in the companion paper [10]. For all samples in range $0.010(6) < x < 0.089(1)$, evidence for a local moment on titanium is found from neutron diffraction and d.c. magnetic susceptibility. Long range antiferromagnetic order obtains for $x \leq 0.071(10)$, while short range order is seen within the range $0.074(9) \leq x < 0.089(1)$. The critical temperatures or spin freezing temperatures lie within the temperature range for which VRH is seen to dominate the conduction process. In any case, there is no evidence for the co-existence of AF order and metallic transport behavior, in contrast to the report on the closely related $Nd_{1-x}Ca_xTiO_3$ system [13].

**IV.B. Metallic samples**

This section focuses on the behavior of the samples which show metallic behavior over a significant temperature range, namely, those with doping levels within the bounds $0.074(9) \leq x \leq 0.203(10)$. These are found in regions II and III on the phase diagram in Figure 2. In this section the main issues to be addressed are the sign of the charge carriers in the metallic regime, the unusual upturn in the resistivity seen at low temperatures in many of the samples which are metallic at high temperatures and the role of electron correlation in the metallic state, extracted by analysis within the Fermi liquid theory.

### IV.B.I. *n*-type metals

From the thermopower data of Figure 6a, all of the metals in this system are *n*-type. The thermopower or Seebeck coefficient of a metal is proportional to the temperature and varies inversely with the carrier density. Within the nearly free electron model, S ~ $m^*n^{-2/3}T$, where $m^*$ is the effective mass and $n$ is the carrier density. It is unlikely that such a simple model will hold quantitatively for a strongly correlated system such as $Nd_{1-x}TiO_3$, but the data of Figure 10 show the expected linear temperature dependence and the increase of S with decreasing $n$ (increasing $x$). Note that the $T^1$ law holds out to ~ 200 K.

### IV.B.II. Upturns in the resistivity at low temperatures

As evident from the ρ(T) results of Figure 5 for the doping range 0.095(8) ≤ $x$ ≤ 0.125(4) NTO, while metallic behavior is observed for most of the temperature range, a clear upturn is seen at low temperatures. Similar upturns appear for higher doping levels from $x$ = 0.145(10) to 0.203(10); results of two compositions are shown in Figure 11 and the data have been obtained from polycrystalline samples. The resistivity rise at low temperatures has been previously seen for $x$ = 0.15 and 0.20 metallic samples in $Nd_{1-x}Ca_xTiO_3$; for this system, the upturns were argued to originate from the localization effects associated with the remanent moments found in samples through susceptibility measurements [13]. Another published report on $La_{0.97}Ti_{0.75}V_{0.25}O_3$ has linked the low temperature resistivity upturns with grain boundary effects [22]. In addition, for the $CaRu_{1-x}Mn_xO_3$ system, the low temperature scattering was ascribed to the Kondo effect due to the local moment $Mn^{4+}$ ions [23].

### IV.B.II.A. Grain boundary scattering

The issue of grain boundary scattering can be addressed by comparison of data obtained on polycrystalline and single crystalline samples of similar compositions. As described earlier, single crystal samples of $Nd_{1-x}TiO_3$ were grown by both Bridgeman and float-zone methods. The electrical resistivity data for few pieces of $x$ ~ 0.15 and 0.20 crystals, cut from the crystal boules, are shown in Figures 12 and 13 as examples. For both crystals, there is only a weak concentration gradient along the crystal lengths, determined by x-ray diffraction. Note the absence of low temperature upturns. Thus, the upturns seen in polycrystalline samples in Figure 11 are likely due to grain boundary scattering effects for these compositions which are fairly deep into the metallic regime, region III in Figure 2.

A different situation prevails for crystals with $x \sim 0.10$, near the MIT1 (Figure 13), where a distinct low temperature upturn is seen for single crystal samples. These were grown by the Bridgeman method and exhibited a somewhat steeper concentration gradient along the axis of the crystal boule. This observation suggests an intrinsic origin to the upturn effect and led to a study of the magnetoresistance of materials with compositions near $x \sim 0.10$.

### IV.B.II.B. Magnetoresistivity

The magnetoresistance of three NTO samples which span the Mott transition, was investigated by comparing the resistivity measured in zero field and an 8 T applied field from 2 K to 300 K (Figure 14a). It is clear that a weak negative magnetoresistance is present with values at 2 K of -1.33%, -1.22% and -1.18% for $x = 0.079(2)$, 0.098(10) and 0.112(4), respectively. From the inset in Figure 14a, the onset temperature for this effect is $\sim 18$ K for $x = 0.079(2)$ which decreases with increasing $x$ to $\sim 12$ K for $x = 0.098(10)$ and $\sim 8$ K for $x = 0.112(4)$. At $x = 0.14$ value, the magnetoresistance effect is expected to disappear (Figure 12). Figure 14b shows the resistivity versus magnetic field for $x = 0.079(2)$. The sample was first zero-field cooled from room temperature to 2 K, and then the field was increased to 9 T and the resistance measured. Sweeping the field back to 0 T produces an increase in the magnitude of the resistivity back to the original value. As seen, there is a smooth decrease of $\rho(H)$ over the field range with an evidence of a weak hysteresis.

Concerning the origin of the weak magnetoresistance effect, it is first important to note that the magnetic properties of the three compositions involved are quite different. In our companion paper on the magnetic consequences of NTO hole doping, it is shown that there exist local moments on both the $Ti^{3+}$ and $Nd^{3+}$ sites in the $x = 0.079(2)$ sample which are involved in short range ordering with a spin freezing temperature of $\sim 38$ K, well above the onset temperature for magnetoresistance [10]. For the other two samples, only the $Nd^{3+}$ sites have a local moment due to the 4f electrons and there is no sign of $Ti^{3+}$ short range order down to 2 K [10]. It thus, appears that any involvement of the remanent magnetization due to titanium moments in the observed magnetoresistance is unlikely, implicating by default the $Nd^{3+}$ moments. The fact that the magnetoresistivity onset temperature decreases systematically with $Nd^{3+}$ concentration supports this conclusion. It is known from previous heat capacity studies on the $Nd_{1-x}TiO_3$ system that the $Nd^{3+}$ moments order near 1 K for doping levels within the range which exhibit the magnetoresistance effect and the ordering was seen in the heat capacity out to much higher temperatures, due presumably to short range ordering [8]. Thus, it is proposed the conduction electrons which are mainly associated with titanium are scattered by the local $Nd^{3+}$ moments via an indirect interaction, probably, RKKY [24]. The application of a magnetic field would orientate the $Nd^{3+}$ moments and decrease the spin disorder component to the scattering, thus altering localization lengths and enhancing the electron mobility, resulting in diminished resistivity.

### IV.B.III. Fermi-liquid Behavior

Experimental evidence for Fermi-liquid (FL) behavior is generally taken as the occurrence of a $T^2$ term in the resistivity, i.e. $\rho(T) = \rho_o + AT^2$, ($\rho_o$: residual resistivity) and linear term in specific heat as $C = \gamma T$. According to the Kadowaki-Woods (KW) relationship, it is shown that $A\gamma^{-2} = 10^{-5}$ ($\mu\Omega$cm mol$^2$K$^2$) mJ$^{-2}$ [25]. The KW relation is $A\gamma^{-2} = n^{-4/3}$ (n: electron density) and it is independent of the effective mass of quasiparticles, $m^*$. This relation is a signature of electron-electron scattering in strongly correlated metallic systems. For the metallic $Nd_{1-x}TiO_3$ samples studied here, the A values have been extracted from the resistivity fits to the above FL expression. As illustrated in Figures 15 to 17, a $T^2$ dependence is revealed in $0.074(9) \leq x \leq 0.196(10)$, in the metallic regions. As evident, similar slopes are found using low temperature data for the single crystal samples in Figure 16 for $x \sim 0.091$ to 0.099, as found for the slightly smaller x values 0.074 to 0.095 from high temperature fits in Figure 15.

In Figure 18, the extracted A coefficients from the FL resistivity fits are plotted as a function of $Nd_{1-x}TiO_3$ doping level. The A coefficients are noticeably enhanced in the vicinity of first metal-to-insulator transition. While this general effect has been seen before in hole-doped titanates and is expected from the arguments of Brinkman and Rice [22], the values attained in the $Nd_{1-x}TiO_3$ system are unprecedented in titanate perovskites. While in previous work [8], only one sample showed an enhanced *A* coefficient, the present study (Figure 7) illustrates that this is a general phenomenon as MIT1 is approached from the metallic regime for this correlated system. Since the KW correlation holds approximately in semi-heavy fermions for $A > 10^{-3}$ $\mu\Omega$cm K$^{-2}$, γ values can be extracted for the highest correlation measured, for example $A \cong 0.03$ $\mu\Omega$cm K$^{-2}$ corresponds to $\gamma \cong 50$ mJ mol$^{-1}$K$^{-2}$. This argument depends on the universality of the KW relationship which has been found to hold well for the closely related $La_{1-x}Sr_xTiO_3$ system where direct measurements of both *A* and γ have been reported [26].

It is difficult to experimentally confirm the apparent γ values and by inference, the large carrier mass enhancements ($\gamma \alpha m^*n^{1/3}$) for compositions near the metal-to-insulator boundaries in $Nd_{1-x}TiO_3$. The direct measurement of γ in heat capacity in this system is hampered by the magnetic ordering of the $Nd^{3+}$ moments near 1 K and the large, associated anomaly in the specific heat which extends to at least 6 K. In a previous study, γ was roughly extracted by fitting heat capacity $Nd_{1-x}TiO_3$ data from 6 K to 10 K, reporting $\gamma = 20.4(8)$ mJ mol$^{-1}$K$^{-2}$ for $Nd_{0.90}TiO_3$ and $\gamma = 11.3(9)$ mJ mol$^{-1}$K$^{-2}$ for $Nd_{0.80}TiO_3$ [8]. For the similar $La_{0.9}Ba_{0.1}TiO_3$, $La_{1.7}Sr_{0.3}CuO_4$ and $Sr_{0.05}La_{0.95}TiO_3$ materials, the maximum observed γ's are reported to be 9.7, ~ 7 and ~ 16 mJ mol$^{-1}$K$^{-2}$, respectively [27, 28, 26]. The largest γ value of ~140 mJ mol$^{-1}$K$^{-2}$ has been reported for the semiconducting $Y_{0.8}Ca_{0.2}TiO_3$ system [26]. This is a curious result which bears re-investigation. $Y_{0.8}Ca_{0.2}TiO_3$ is reported to be an insulator and thus, one would expect γ = 0 as there is no Fermi surface. Thus, the inferred gamma value (from the KW relationship) for samples near $Nd_{0.92}TiO_3$ of ~50 mJ mol$^{-1}$K$^{-2}$ appear to be the largest yet reported for a metallic titanate perovskite.

## V. CONCLUSIONS

The transport property investigations presented here for $Nd_{1-x}TiO_3$ is combined with magnetic properties study (our report in [10]) in form of a phase diagram and compared with reported results on related systems such as $La_{1-x}Sr_xTiO_3$, $La_{1-x}TiO_3$, $Y_{1-x}Ca_xTiO_3$ and $Nd_{1-x}Ca_xTiO_3$. Initially, $NdTiO_3$ was chosen as the focus of this study for several reasons. It is clearly established, from optical and magnetic data, as an antiferromagnetic Mott-Hubbard insulator with a Mott gap of ~ 0.8 eV. Large correlation effects were expected relative to $LaTiO_3$ due to more acute Ti – O – Ti bond angles and narrower lower Mott-Hubbard bandwidth. Also, a convenient hole-doping mechanism was available through the creation of Nd site vacancies, $Nd_{1-x}TiO_3$.

Fine control of the doping levels which span the Mott transition have revealed a lot more detail, not reported previously, which are summarized in the phase diagram in Figure 19. For hole-doping $0.010(6) \leq x \leq 0.071(10)$, the system is insulating or semiconducting with predominantly *p*-type carriers. Activation energies ($E_a$) decrease significantly from ~ 0.18 eV to 0.03 eV with doping and must be associated with excitations to mid gap states created by the hole doping and not to the higher lying upper Mott-Hubbard band. Long range antiferromagnetic order occurs with Neél temperatures and ordered $Ti^{3+}$ moments which decrease from 88 K and 0.45 $\mu_B$, in $x = 0.010(6)$, to 60 K and 0.31 $\mu_B$ in $x = 0.071(10)$.

The $0.071(10) \leq x \leq 0.095(8)$ compositions show apparent transitions in resistivity from semiconducting behavior at low temperatures to a metal-like behavior above ~ 175 K. The sign of the carriers for these samples is predominantly *n*-type. Within the semiconducting temperature range, there is unmistakable evidence for disorder induced localization in the form of correlated electron variable range hopping between 27 K and 63 K which gives way to activated hopping with very small $E_a$ of order < 0.02 eV which extrapolates to zero at $x = 0.089(1)$. Within this composition range, we have found drastic changes in the magnetic properties. The ordered $Ti^{3+}$ moment falls precipitously from 0.25(4) $\mu_B$ to values below the detection limit of ~ 0.06(10) $\mu_B$ from $x = 0.071(10)$ to $x = 0.074(9)$. Long-range antiferromagnetic order gives way to short-range order with spin freezing temperatures between 30 K and 40 K for $0.074(9) \leq x \leq 0.089(1)$. For no temperature range does AF order of any kind co-exist with metallic behavior. Note that the disappearance of antiferromagnetic long-range order and an ordered $Ti^{3+}$ moment occurs at a significantly lower doping level, $x = 0.071(10)$, than the disappearance of the activation energy, $x = 0.089(1)$, and the onset of the metallic paramagnetic state. The compositions within this range span the Anderson-Mott transition.

Finally, for $0.095(8) \leq x \leq 0.243(10)$, Fermi-liquid metallic behavior is seen at all temperatures. Slight upturns at low temperatures can be ascribed to grain boundary scattering for $x > 0.14$ but a weak, 1%, negative magnetoresistivity effect seen for $0.079(2) \leq x \leq 0.112(4)$, below 15 K, implicates a magnetic scattering interaction with the $Nd^{3+}$ moments. Noticeably large $A$ Fermi-liquid $T^2$-coefficients occur over the range $0.074(9) < x < 0.095(8)$, suggestive of a large mass enhancement as the Anderson-Mott transition is traversed. Initially and for $x = 0.095(8)$, the insulator-to-metal transition appears to be driven by the overlap between the lower Mott-Hubbard band and the mid-gap states rather than the collapse of the Mott-Hubbard gap. This is supported by our optical conductivity data [4]. It is not yet known if the Mott-Hubbard gap does indeed collapse for doping levels with $x > 0.095(8)$ and additional optical studies are underway.

In conclusion, it is important to compare the results of $Nd_{1-x}TiO_3$ phase diagram, with others obtained previously on related systems. The $Nd_{1-x}Ca_xTiO_3$ series [7] is most closely related to the present study. For these Ca-doped materials, the onset of the paramagnetic metallic state is reported for $x = 0.20$ which corresponds to a 20% hole ($Ti^{4+}$) concentration. For the $Nd_{1-x}TiO_3$ system, the equivalent doping level is $x = 0.098$ or a 29% doping level. This suggests a higher level of correlation for the vacancy doped materials. In fact the relevant hole concentration for the $Sm_{1-x}Ca_xTiO_3$ series is reported at about 28% and that for $Y_{1-x}Ca_xTiO_3$, that is the most strongly correlated titanate system, is 34%. The 29% $Nd_{1-x}TiO_3$ critical doping level is much greater than that for the much less correlated $La_{1-x}Sr_xTiO_3$ (~ 5%) and greater than $La_{1-x}TiO_3$ (24%). Also, in the $R_{1-x}Ca_xTiO_3$ systems (R = La, Pr, Nd, Sm), the Mott transition is associated with a discrete doping level, not a range, as we found here. As well, the first insulator-to-metal transition is reported at a lower doping level than the disappearance of antiferromagnetic order, just the opposite as found for the $Nd_{1-x}TiO_3$ system.

## Acknowledgements

We thank the Natural Sciences and Engineering Research Council of Canada for support in the form of a pre-doctoral fellowship for A. S. Sefat and a Discovery Grant to

J. E. Greedan, G. M. Luke, M. Niewczas. The support from CIAR Quantum Materials program to G.M. Luke is appreciated. In addition, the authors acknowledge the financial support by the MMO-EMK Research Grant.

**Table 1**: Summary of the ρ(T) range fitted to Arrhenius law and the activation energies, $\Delta_{act}$, obtained for $Nd_{1-x}TiO_3$ samples.

| $x$ | NTO formula | Range fitted (K) | $\Delta_{act}$ (eV) | $\Delta_{act}$ (K) |
|---|---|---|---|---|
| 0.010(6) | $Nd_{0.990(6)}TiO_3$ | 90 - 340 | 0.084 | 976 |
| 0.046(10) | $Nd_{0.954(10)}TiO_3$ | 56 - 140 | 0.048 | 553 |
| 0.057(11) | $Nd_{0.943(11)}TiO_3$ | | 0.020 | 233 |
| 0.071(10) | $Nd_{0.929(10)}TiO_3$ | | 0.021 | 243 |
| 0.074(9) | $Nd_{0.926(9)}TiO_3$ | ~ 63 - 145 | 0.010 | 114 |
| 0.079(2) | $Nd_{0.921(2)}TiO_3$ | | 0.007 | 85 |
| 0.089(1) | $Nd_{0.911(1)}TiO_3$ | | 0.005 | 59 |
| 0.095(8) | $Nd_{0.905(8)}TiO_3$ | | 0.002 | 25 |

**Figure 1:** Room temperature optical conductivity for $Nd_{0.981(6)}TiO_3$ with $x = 0.019(6)$. Dashed lines show the linear extrapolation of the edge of the Mott-Hubbard gap ($E_g$) and charge-transfer transition ($\Delta$).

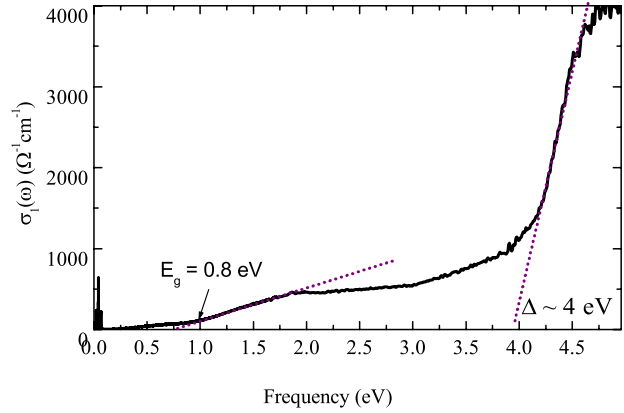

**Figure 2:** The electronic model for the $Nd_{1-x}TiO_3$ (NTO) solid solution; MITs are the metal-insulator transitions. The NTO composition range under investigation is indicated by the arrow. $E_f$ is the Fermi energy, $\Delta_{act}$ is the conduction activation energy, $E_g$ is the Mott-Hubbard gap and $\Delta$ is the charge-transfer energy.

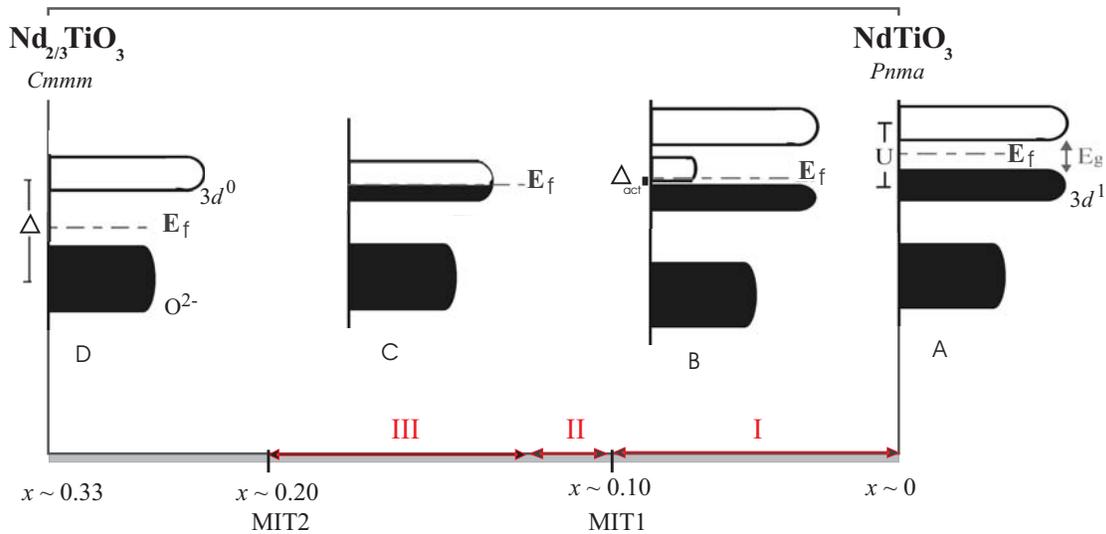

**Figure 3:** The view of *Cmmm* structure for $Nd_{0.70}TiO_3$ (a) and *Pnma* structure for $Nd_{0.98}TiO_3$ (b). In *Cmmm* model, there are two $Nd^{3+}$ sites and $z = 1/2$ is partially occupied.

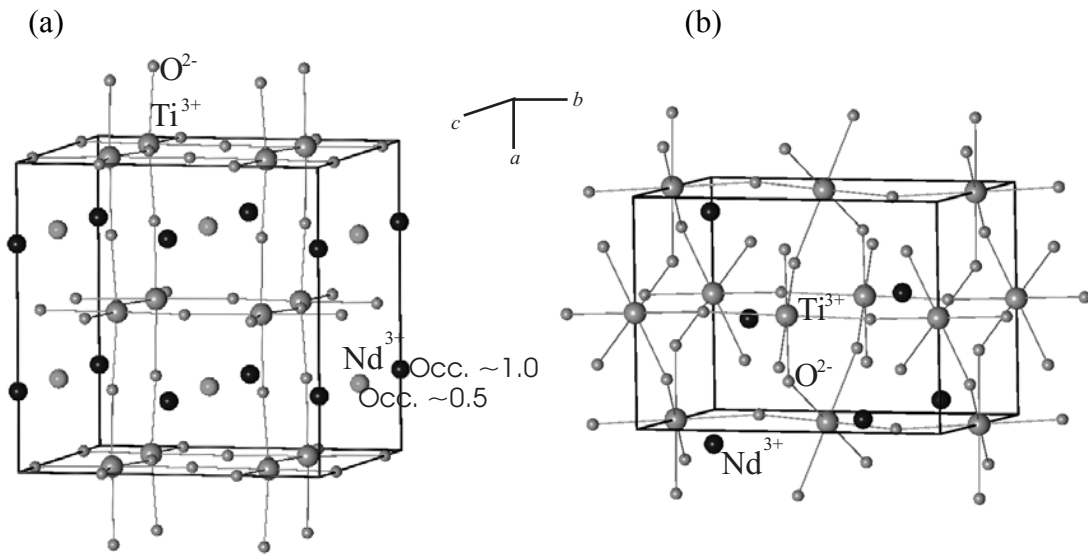

**Figure 4:** The (a) Ti – O – Ti bond angles and (b) Ti – O bond distances as a function of neodymium vacancies, $x$, in $Nd_{1-x}TiO_3$ solid solution.

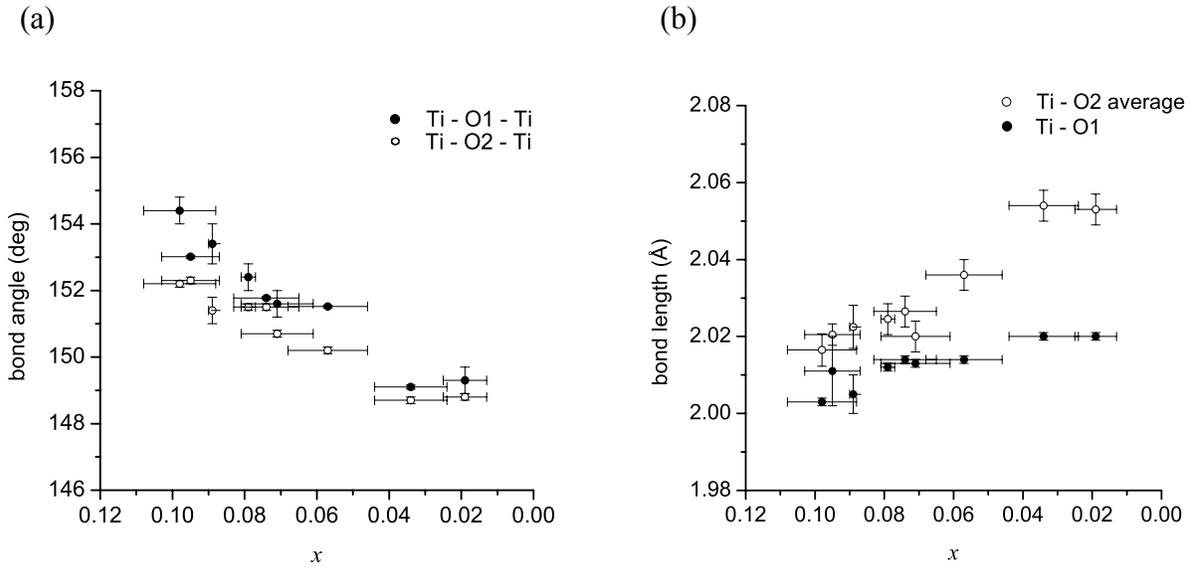

**Figure 5:** Temperature dependences of the electrical resistivity ρ(T), versus *x*, for $Nd_{1-x}TiO_3$ samples spanning the first metal-to-insulator transition.

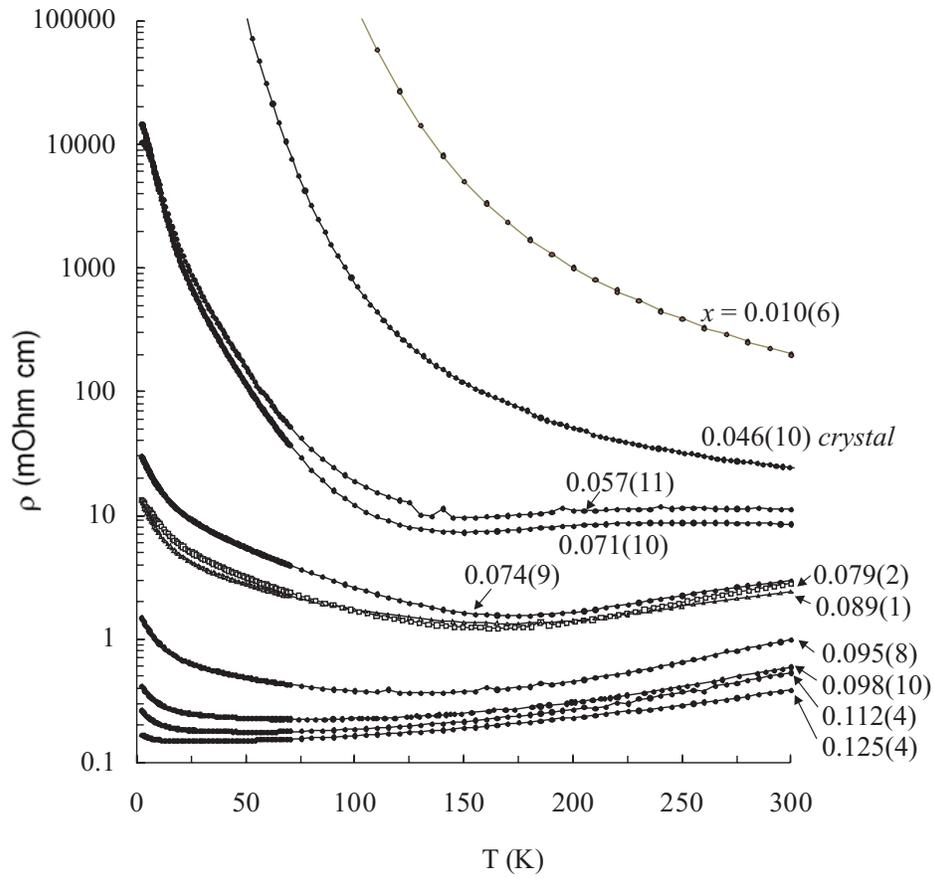

**Figure 6:** Temperature dependence of the thermoelectric power S(T), versus *x*, for (a) $Nd_{1-x}TiO_3$ samples; (b) is the reprinted figure for $La_{1-x}Sr_xTiO_3$, taken with permission from reference [9].

(a)
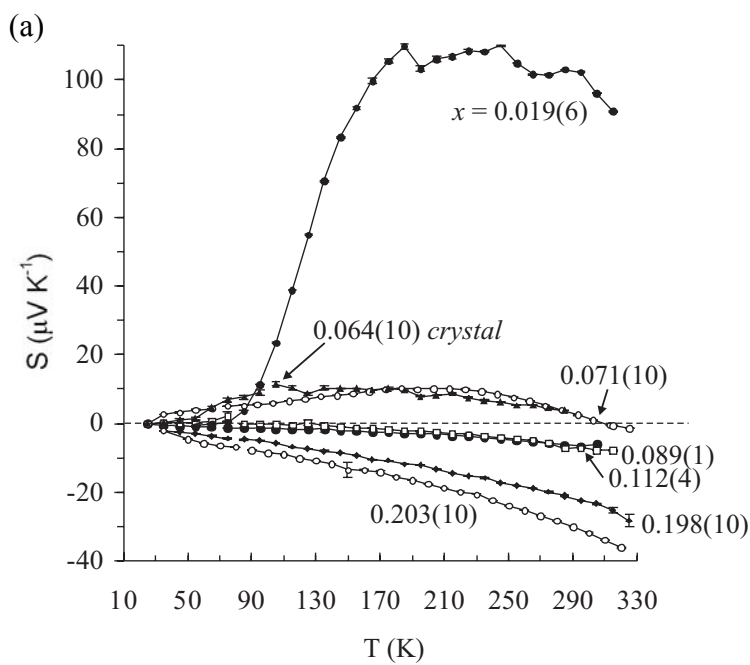

(b)
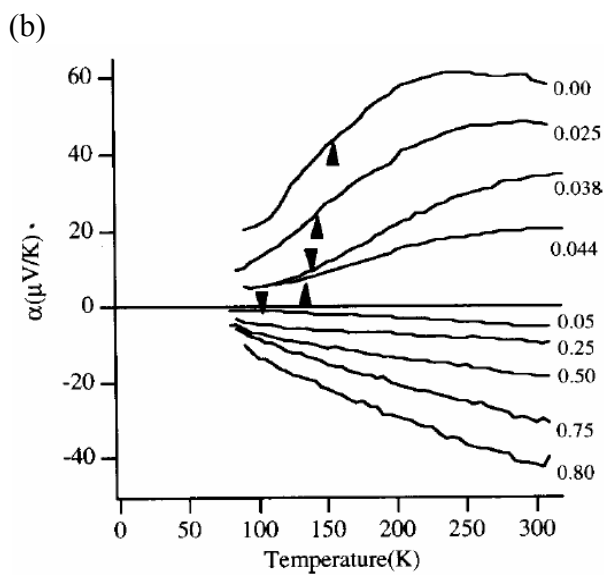

**Figure 7:** The high temperature $\rho(T)$ Arrhenius plots for (a) $Nd_{1-x}TiO_3$ and (b) the band gap ($2\Delta_{act}$) for activated hopping conduction and the corresponding activation energy ($\Delta_{act}/k$) versus $x$.

(a)
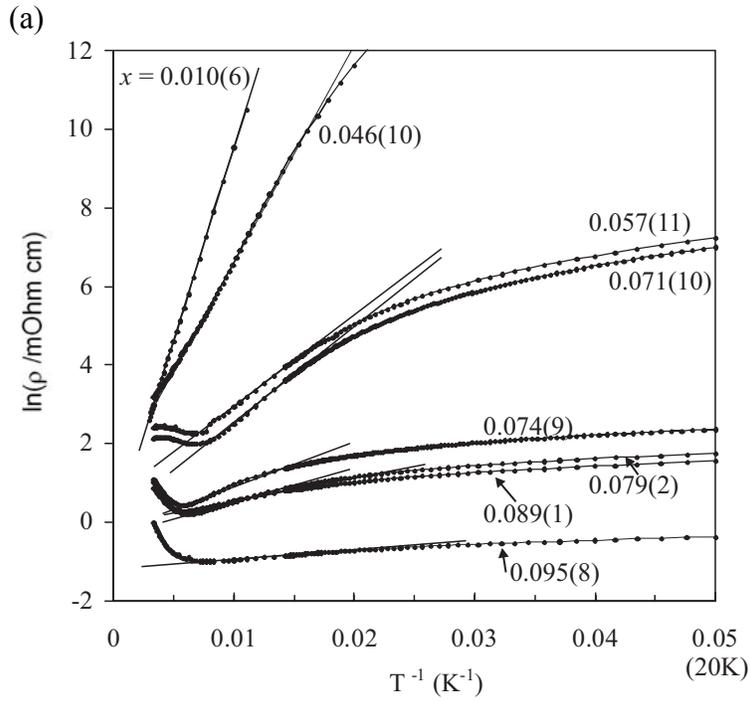

(b)
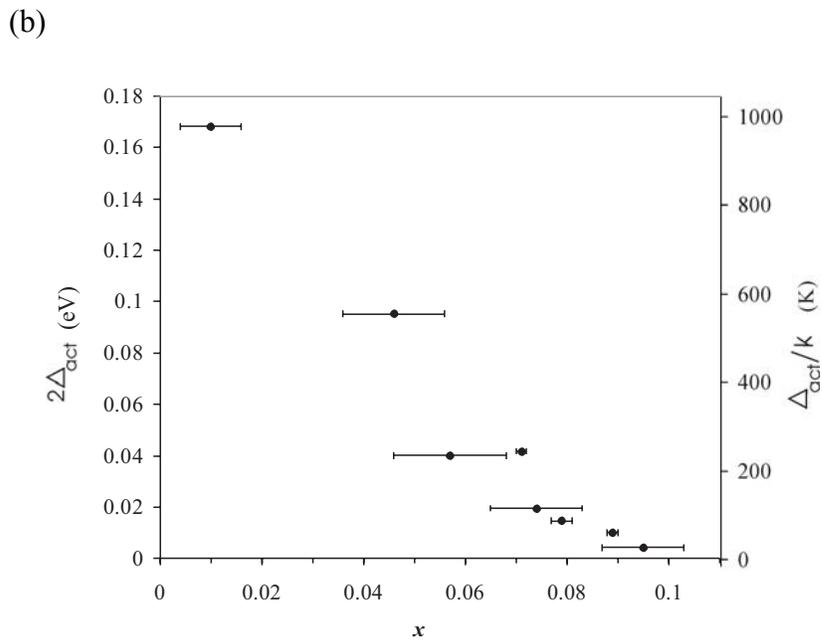

**Figure 8**: The log plot of effective energy (E) versus temperature derived from the resistivity data for various $Nd_{1-x}TiO_3$ samples in the range of 2 K – 158 K. The fits from 27 K to 63 K are shown in (a); the slope of line is the y hopping index. The summary plot of y exponent versus $x$ is shown in (b).

(a)
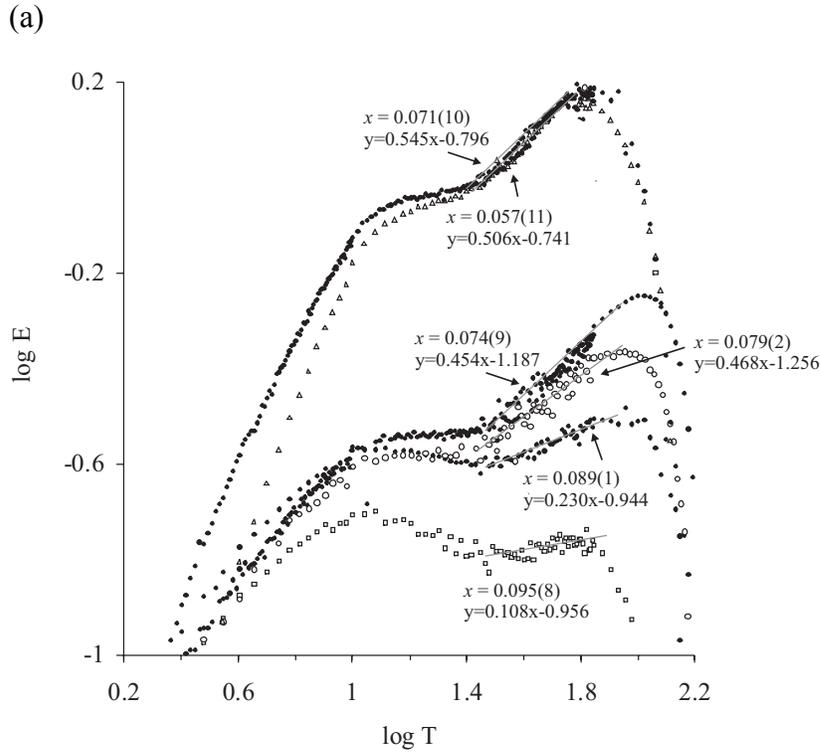

(b)
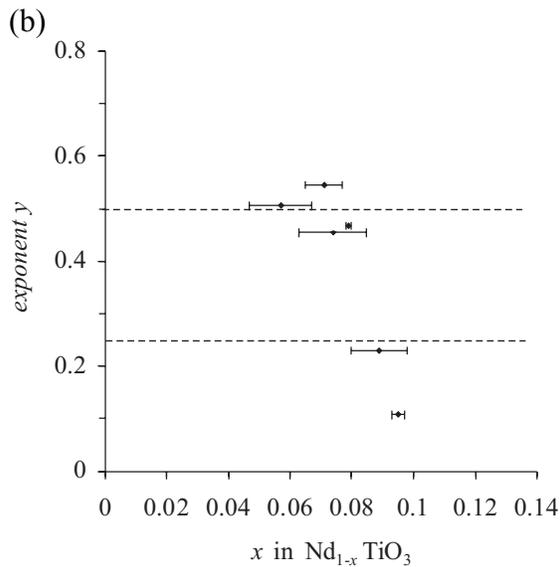

**Figure 9**: Temperature dependences of the electrical resistivity ρ(T) for $Nd_{0.921(2)}TiO_3$, with $x = 0.079(2)$.

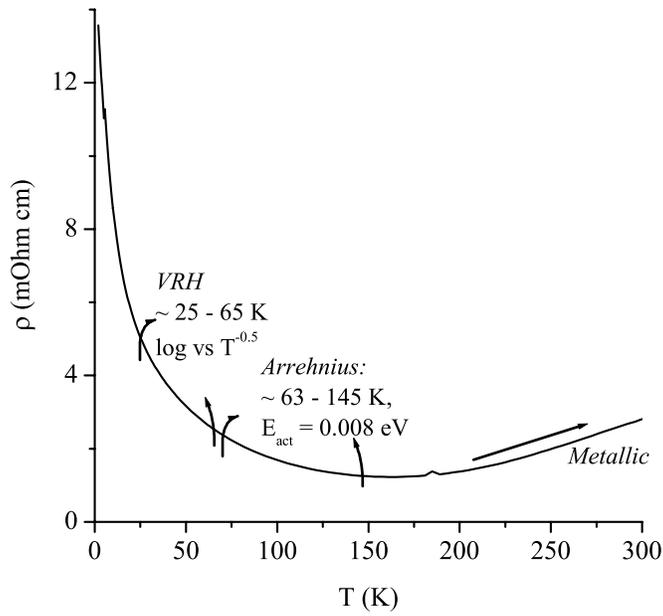

**Figure 10:** Evidence for metallic behavior in the region of ~ 25 – 200 K in the metallic compositions of NTO.

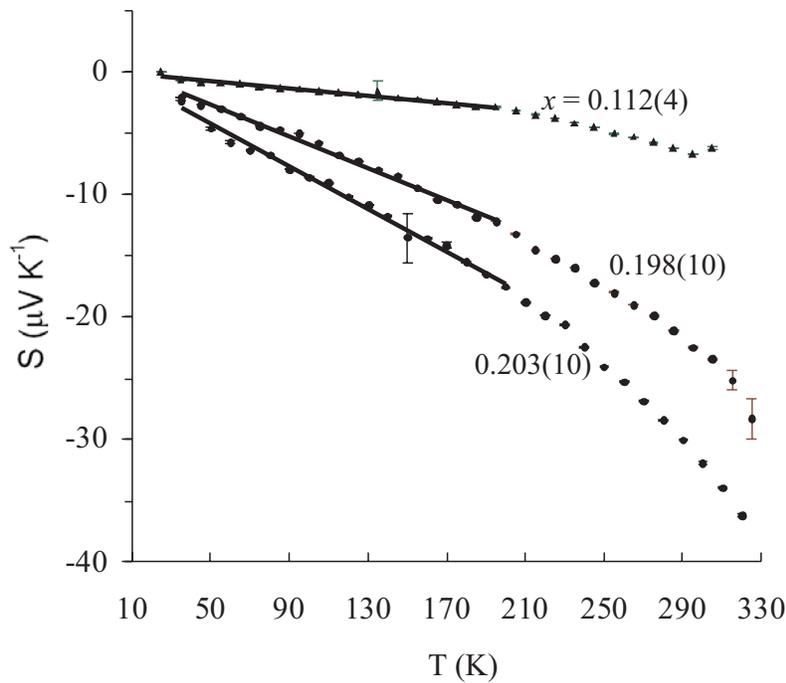

**Figure 11:** The ρ(T) for two $Nd_{1-x}TiO_3$ polycrystalline samples. Note the presence of upturns at low temperatures.

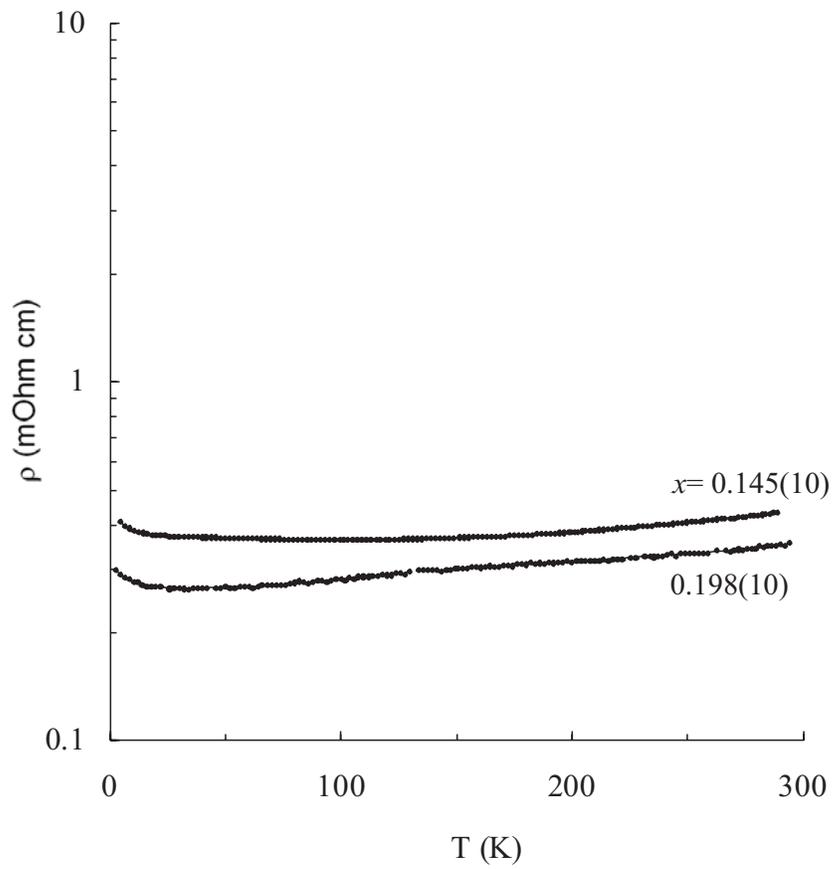

**Figure 12:** The electronic resistivity vs temperature for single crystal pieces with $x \sim$ 0.15 and 0.20 compositions in $Nd_{1-x}TiO_3$. Note the absence of upturns at low temperatures.

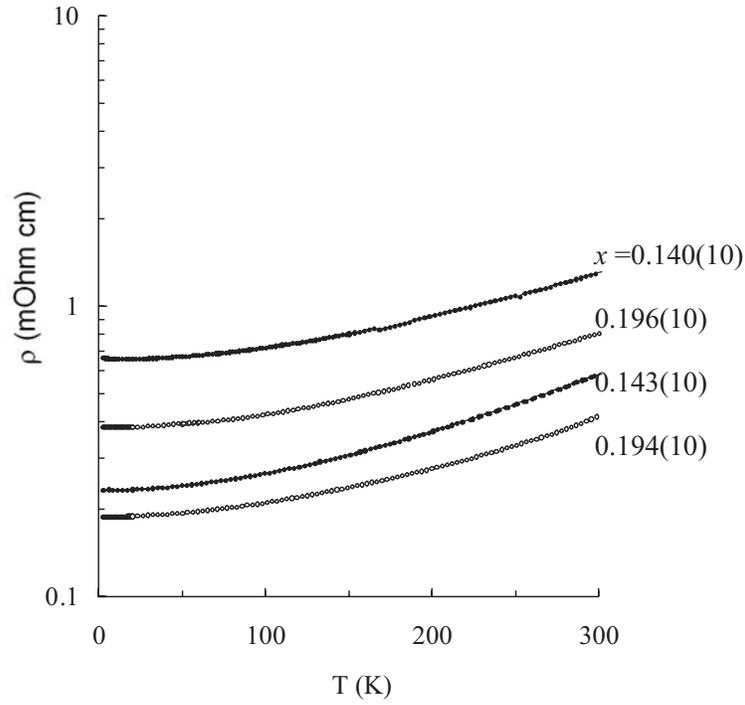

**Figure 13:** Temperature dependence of resistivity for various pieces of $x \sim 0.10$ $Nd_{1-x}TiO_3$ crystal. Note the persistence of the low temperature upturns in the single crystal samples.

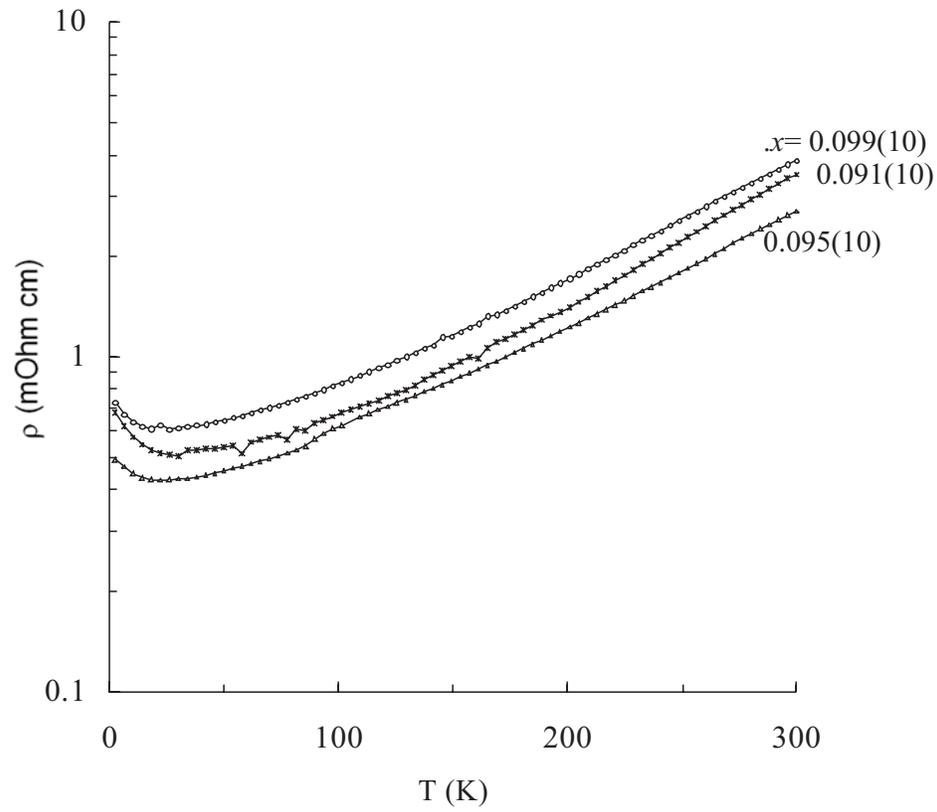

**Figure 14:** a) The $\rho(T)$ results at 0 T and 8 T for $Nd_{0.921(2)}TiO_3$, $Nd_{0.902(10)}TiO_3$ and $Nd_{0.888(4)}TiO_3$ samples; b) the $\rho(H)$ at 2 K for $Nd_{0.921(2)}TiO_3$, with $x = 0.079(2)$. The inset in (a) is the zoomed low temperature data, showing the field divergence more clearly.

(a)
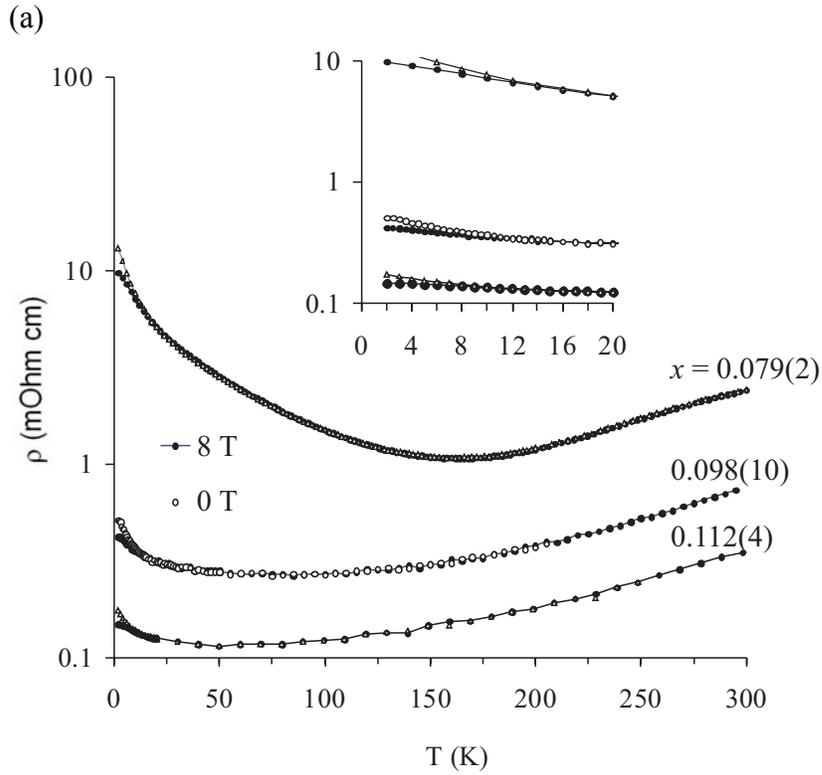

(b)
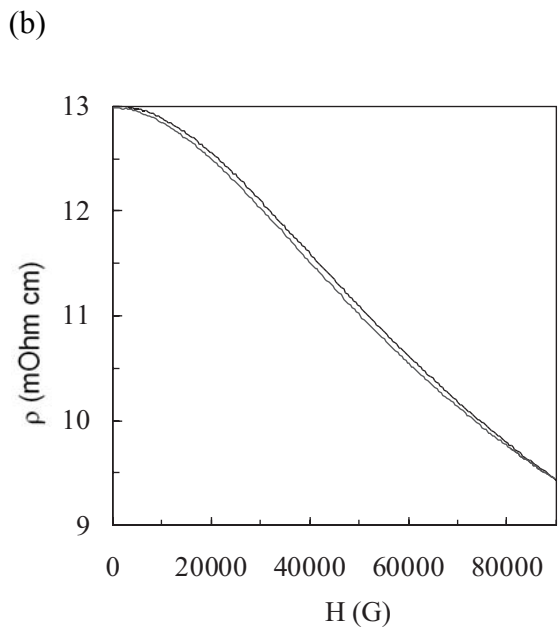

**Figure 15:** The Fermi-liquid behavior illustrated from $\rho$ vs $T^2$ graphs for the $x$ polycrystalline samples at the first metal-to-insulator transition in $Nd_{1-x}TiO_3$.

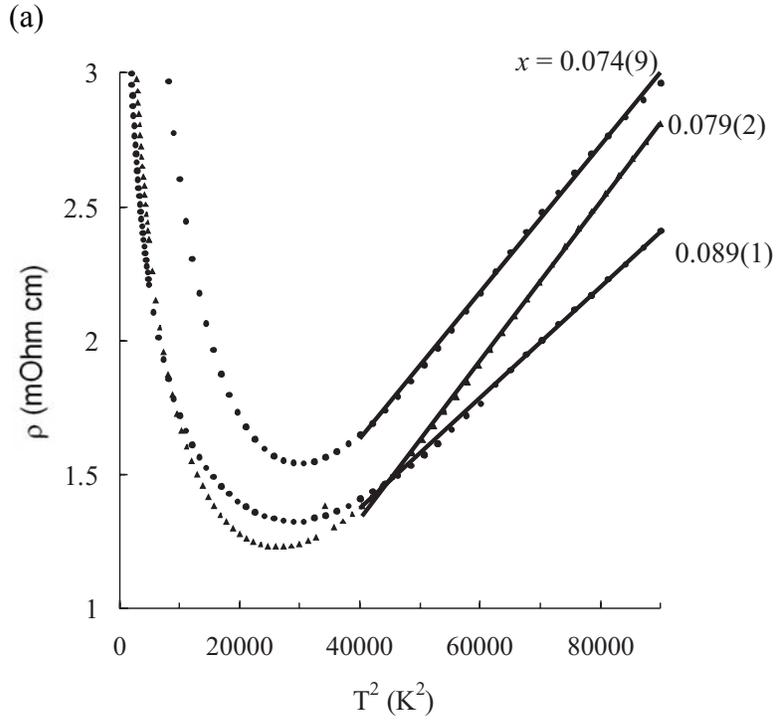

(a)

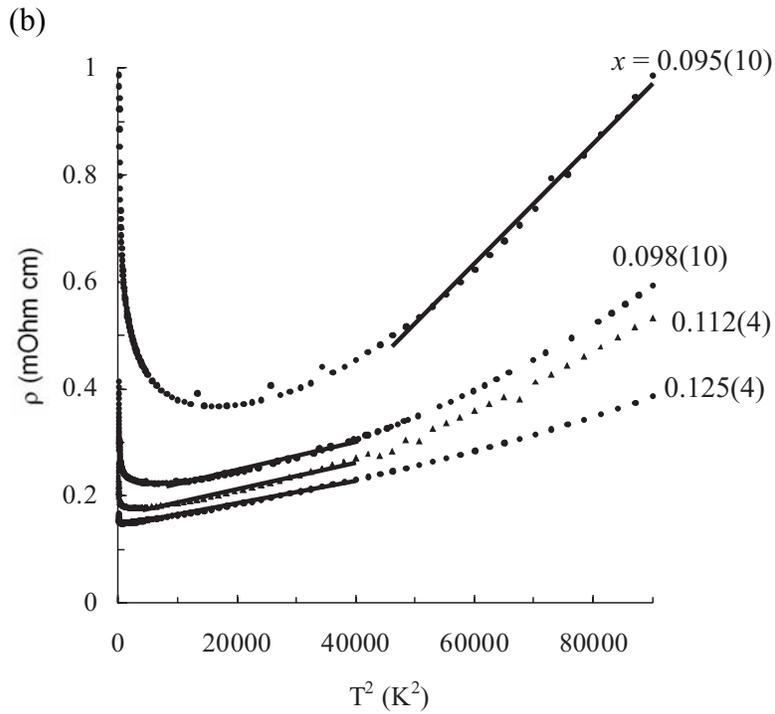

(b)

**Figure 16:** The Fermi-liquid behavior illustrated from $\rho$ vs $T^2$ graphs for the single crystalline metallic $x \sim 0.10$ members of the $Nd_{1-x}TiO_3$ series; $x$ values are noted on the graphs.

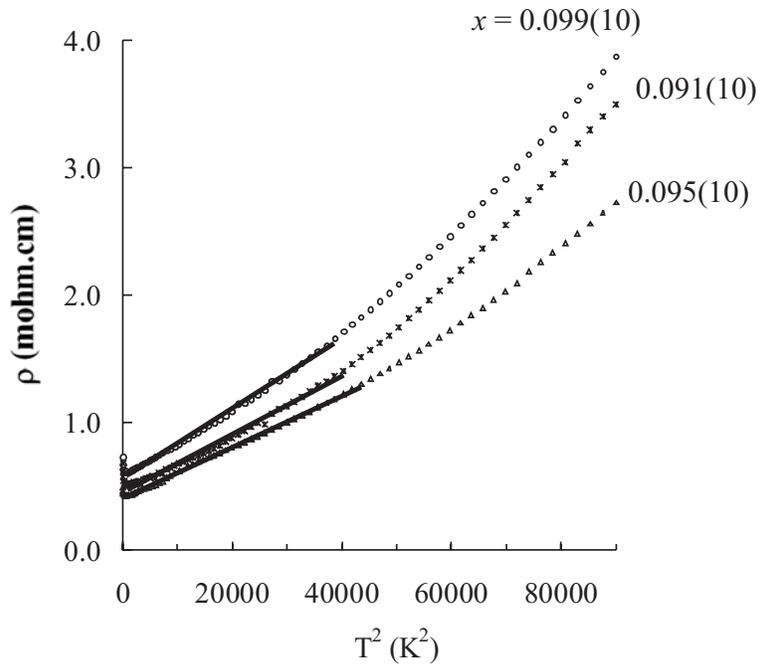

**Figure 17:** The Fermi-liquid behavior illustrated from $\rho$ vs $T^2$ graphs for the crystalline metallic $x \sim 0.15$ and $0.20$ members of the $Nd_{1-x}TiO_3$ series; $x$ values are noted on the graphs.

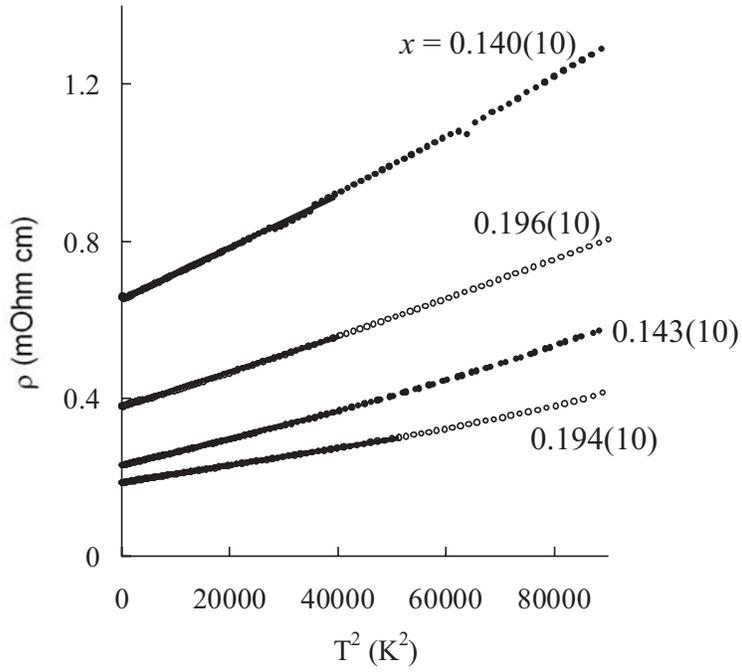

**Figure 18:** The extracted *A* coefficients from Femi Liquid fits to ρ(T) data in Nd$_{1-x}$TiO$_3$ samples. Note that the *A* coefficients are noticeably enhanced towards the first metal-to-insulator transition.

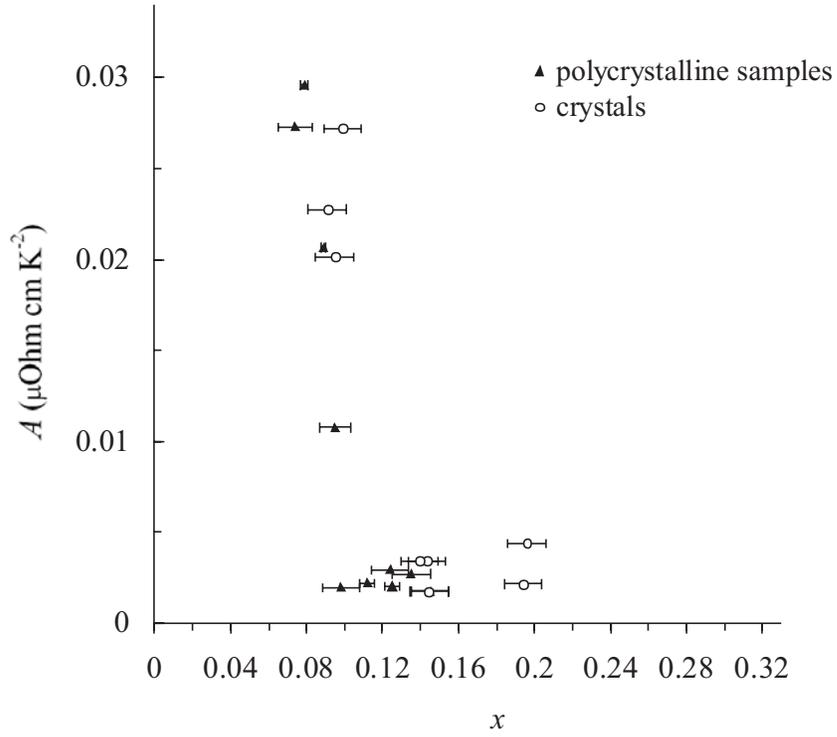

**Figure 19**: The phase digram of $Nd_{1-x}TiO_3$ with $0.33 \leq x \leq 0$. 'PM I' is the paramagnetic insulator; 'PM M' indicates the paramagnetic and Fermi-liquid metal phase; 'SR AFI' indicates the short-range-ordered antiferromagnetic Mott insulator; 'LR AF I' is the long-range-ordered antiferromagnetic Mott insulator. The filled and empty symbols show $T_{order}$ for LR and SR AF compositions, respectively.

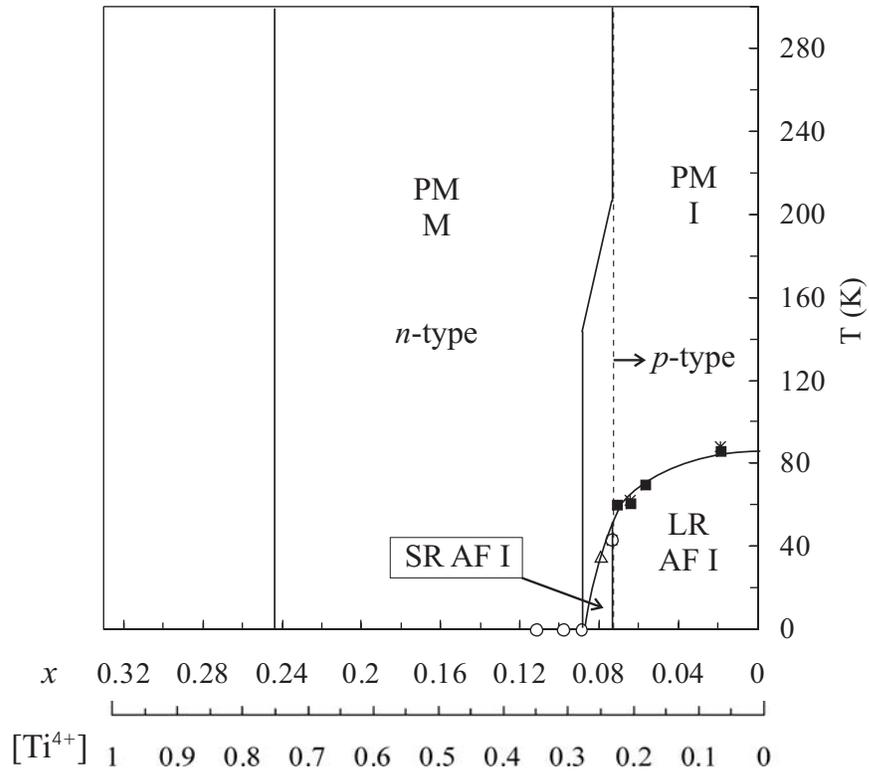